 \UseRawInputEncoding
\documentclass[aps,prl,twocolumn,groupedaddress]{revtex4}
\usepackage{graphicx}

\begin{document}

\title{The pseudogap in hole-doped cuprates: possible insights from the Kondo effect}


\author{J. R. Cooper}
\affiliation{ Cavendish Laboratory, Department of Physics, University of Cambridge, J.J. Thomson Avenue, CB3 0HE, United Kingdom}


\date{\today}


\begin{abstract}
The ``states-non-conserving'' fermion density of states (DOS), deduced from the specific heat of hole-doped cuprates,
 could arise from a Kondo or heavy fermion-like DOS being suppressed by anti-ferromagnetic spin
  fluctuations. The large Fermi surface predicted by band theory and observed experimentally,
is still expected for zero pseudogap, but with an effective mass corresponding to a Kondo temperature
 $T_K\simeq$800~K. A finite pseudogap could divide it into Fermi arcs. Theoretical results for the asymmetric Anderson
  model can account for the experimental Wilson ratio.
\end{abstract}

\pacs{}

\maketitle

One of the  unsolved questions for cuprate superconductors, in addition to  the pairing mechanism and the linear temperature ($T$) dependence of the
 electrical resistivity, is
 the origin of the pseudogap~(PG) in samples   with a certain number of  holes~($p$) per CuO$_2$ unit. The PG lowers the superconducting condensation
  energy~\cite{LoramJPCS1998,LoramJPCS2001},
   the superfluid density, e.g.~\cite{Anukool2009},
 and an important practical property, the irreversibility
field~\cite{Cooper1997} where the electrical resistivity first becomes non-zero, even though the zero-field transition temperature initially stays high.
  Many researchers  focus on a $T^*-p$ ``phase diagram'' and a $T^*(p)$  line, ascribed to the onset of a  PG, that
   goes to zero as $p$ is increased to
   $p_{crit}$~\cite{Keimer2015}.
Careful  experiments  reveal anomalies in  several  physical properties at, or near
   $T^*$~\cite{Keimer2015}, but the values of $T^*$ and the size of the anomalies are probably too small to account for the PG.
For example  the small~\cite{Kokanovic2016} charge density wave (CDW) anomalies for under-doped (UD) samples may be
 linked~\cite{Comin2014} to the
 Fermi arcs, detected by angle-resolved photoemission~(ARPES),
 that are caused by the pre-existing PG. Abrupt changes
  in magnetic anisotropy of YBa$_2$Cu$_3$O$_{6+x}$~(YBCO) crystals defining a nematic $T^*(p)$ line~\cite{Matsuda2017}, that
   also extrapolates to zero at $p_{crit}$,
are also small - typically  0.5$\%$ of the average in-plane spin susceptibility for
 $\Delta T \sim$50~K.
     Here we retain the idea, reviewed along with many experimental facts  in ~\cite{TallonLoram2001},  that the
    PG is best described as a $T$-independent energy scale $E_G$ that varies strongly with $p$. We argue that
     it might be caused  by low energy
 anti-ferromagnetic~($af$) spin fluctuations giving  a deep dip in a Kondo or heavy fermion-like density of states~(DOS) at the Fermi
 energy~($E_F$).

\begin{figure}
\includegraphics[width=75mm,height=75mm]{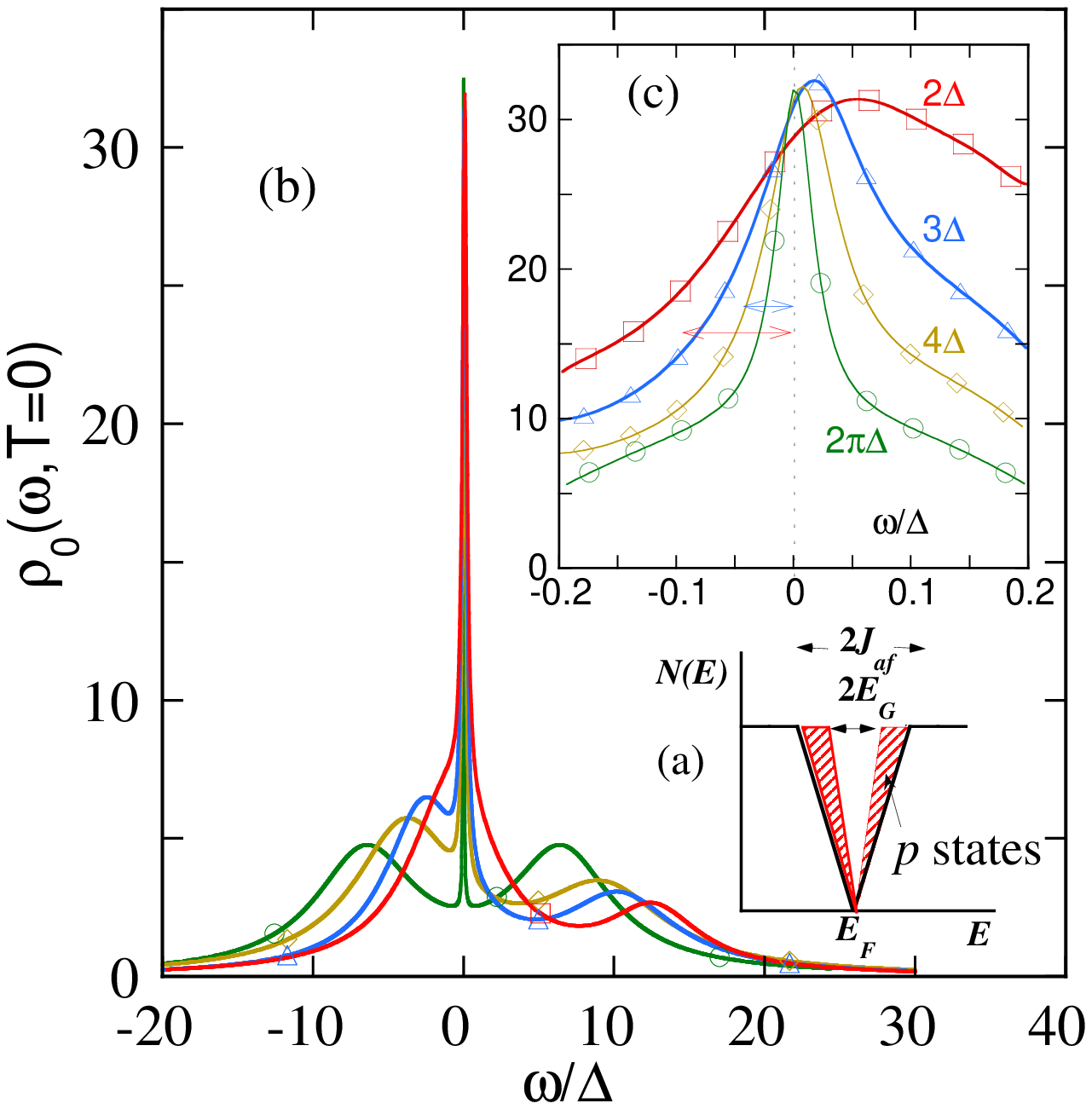}
\caption{Color online. (a) Fermion DOS~\cite{LoramJPCS1998} used
 to rationalize specific heat data. (b) and (c), taken from~\cite{Costi1994} with permission, triple peak structure
  in the spectral density  $\rho_0(\omega,T =0)$ calculated using NRG theory for the single orbital, asymmetric
  Anderson model with $S_{imp}$ =1/2, for
various values of $-E_0/\Delta$ = 2$\pi$, 4, 3 and 2. Here we estimate $-E_0/\Delta = 2.3$ for  all the Cu ions in Bi2212.
 The horizontal
arrows show $T_K/\Delta/2$ for $-E_0/\Delta$= 2 and 3 from~\cite{Costi1994}. Here the factor 2  ensures consistency
between Eqn.~\ref{gamma}~\cite{Wilson1975} and~\cite{Costi1994}, details are given in~\cite{Suppmat}.} \label{Fig1Costi}
\end{figure}

Analysis of the specific heat of YBCO~\cite{Loram1993},
  Bi$_2$Sr$_2$CaCu$_2$O$_{8+x}$~(Bi2212)~\cite{LoramJPCS2001},
 at different oxygen levels~($x$), and La$_{2-y}$Sr$_y$CuO$_4$
~(LSCO)~\cite{LoramJPCS2001}. in terms of a ``states-non-conserving" V-shaped
fermion DOS~\cite{LoramJPCS1998}, centered
 at $E_F$, in an otherwise
  flat band, as sketched in Fig.~\ref{Fig1Costi}(a), gives intriguing insights.
 It has has some deficiencies because (i)
 the large thermoelectric power~(TEP) and its scaling with
$E_G$~\cite{Cooper2000}, require electron-hole asymmetry, (ii) the  detection of quantum oscillations and  Fermi arcs
 implies that there is a small residual DOS at $E_F$ and
 (iii) Y$^{89}$ NMR Knight
shift data for over-doped (OD) Y(Ca)BCO~\cite{Williams1998} show a fall  of up to 20$\%$ in the spin susceptibility,~$\chi_s(T)$, at higher $T$ rather
 than the constant value expected
 for a flat band, there are also small deviations from flat band behavior for OD Bi2212~\cite{LoramJPCS2001}.
 On the other hand  Cu NMR  data for La-doped Bi2201~\cite{Kawasaki2010} seem to be consistent with a flat band.
 They also show that the PG is not affected by fields up to 44~T, and that the residual DOS is much larger in Bi2201.
  The main features of this first-order phenomenological analysis are as follows.

A. $E_G$  falls  linearly with $p$,   as $J_{af}(1-p/p_{crit})$, disappearing
 abruptly at
$p_{crit}$ = 0.17-0.19~\cite{LoramJPCS1998}, where $J_{af}/k_B$ = 1200-1500~K, is similar to
 the  $af$ interaction $J_{af}/k_B$= 1700~K~\cite{Rossat1992}
in the corresponding parent insulator with $p=0$ and $k_B$ is Boltzmann's constant.

 B. Plots of $S_e(T,p)$ $vs.$ $T$, where $S_e(T,p)$ is the
non-phonon, electronic contribution to the entropy,
 are approximately linear and parallel  at higher $T$, e.g.  Fig.~1(b) of ~\cite{LoramJPCS2001},
 and Fig.~5 of~\cite{Loram1993}, or Figs.~1(SM) and 2(SM) of~\cite{Suppmat}.
$S_e(T,p)$  curves   for  $p<p_{crit}$  have lower entropy at all $T$ and
 there is no sign of the entropy ``coming back'' up to
 at least 200~K~ for Bi-2212 and 300~K for YBCO.
 This behavior is consistent with a recent comprehensive NMR paper~\cite{Haase2020}.
 For  fermions with an energy($E$)-dependent DOS
  we expect $S_e(T)$  $\alpha$ $T\chi_s(T)$, because both depend on the number
   of states within the thermal window  $|E-E_F|\lesssim 2k_BT$.  $T\chi_s(T)$ curves
  for  Bi2212  are parallel
   up to 400~K, see Fig.~13(b) of~\cite{LoramJPCS2001} or Fig.~3(SM)~\cite{Suppmat},
    as are plots of $T\chi_s(T)$
 for  Y$_{0.8}$Ca$_{0.2}$Ba$_2$Cu$_3$O$_{6+x}$,  also up to 400~K~\cite{Naqib2009}, after allowing  for an
   $x$-independent  Curie or Curie-Weiss term.  The difference between a states-conserving  CDW- or superconductor-like DOS
and a V-shaped gap
   is shown by the results of model calculations of  $S_e(T)$
    in Fig.~4(SM)~\cite{Suppmat}.

C.  Plots of $S_e(T_0)$ $vs.$ $p$ and $k_BT_0\chi_s(T_0)$ $vs.$ $p$ at
 temperatures $T_0$ show  anomalously large increases
 of $\approx k_B$  and 0.3$\mu_B^2$  per added hole respectively~\cite{LoramJPCS2001}, where $\mu_B$ is the Bohr magneton.
 For $p<p_{crit}$  this is  caused by the $p$-dependence of the PG and is consistent with
 the  Wilson ratio discussed later.
 For $p>p_{crit}$  they peak near
$p=0.25$ before falling again, but this interesting result
 has only been established experimentally for LSCO~\cite{LoramJPCS2001}.

D. $p_{crit}$ seems to be determined by the product of $J_{af}$ and the  DOS at high energy~\cite{LoramJPCS2001}. In contrast
 to points A to C this
 is not expected
in the present  picture
   because here  $J_{af}$ in the parent compound and the DOS for $E\gg E_{G}$ are not related.

    It is widely
    accepted that the Hubbard model describes the basic physics of the cuprates,  e.g.~\cite{Georges2018} and references therein,
 but it has not been solved rigorously at low $T$ and as far as we know does not account for points A to D above.
Similar underlying physics occurs in
 the Friedel-Anderson model~\cite{Anderson1961} and
the Kondo effect~\cite{Gruner1974,Hewson}, where  metals such as Cu or Au are alloyed with
 a small concentration~($c_{imp}$) of magnetic elements~(impurities)  such as
Fe or Mn. For both Kondo alloys and heavy fermion compounds, the electronic specific heat coefficient, $\gamma\equiv dS_e/dT$,
 is strongly enhanced at low $T$, in Wilson's
solution of the Kondo problem~\cite{Wilson1975} it  is given by
 \begin{equation}
    {\gamma_{imp} =c_{imp}\pi^2 k_Bw/(6T_K)}\label{gamma}
 \end{equation}
 where $T_K$ is the Kondo temperature and $w$=0.4128 is the Wilson number.
For impurities with spin 1/2,  and $T \ll T_K$,
 $\chi_s^{imp} = c_{imp}(g\mu_B)^2w/(4k_BT_K)$, where $g \simeq 2$ is the $g-$factor  and $\chi_s^{imp}/\gamma_{imp}=R_W=2R_0$,  in agreement with
  experiments on some Kondo alloys~\cite{Hewson}. Here  $R_0=3\mu_B^2/(\pi^2k_B^2)$ is the Wilson ratio for
   non- or weakly-interacting fermions.

For Bi2212, $\gamma(200)$ varies from 0.86 to 0.98 mJ/g-at/K$^2$ for  0.095 $<p<$ 0.22~\cite{LoramJPCS2001},
and taking $c_{imp}$ to be the concentration of Cu atoms, Eqn.~\ref{gamma} gives a first estimate for $T_K\simeq$800~K.
  We note that such a single ion Kondo model
works well for  the heavy fermion compound CePb$_3$  doped with La~\cite{CePb}. For the same range of $p$,
    $\chi_s(200)$ = 2.18 to 2.38 10$^{-4}$ emu/mole Bi2212, see footnote~\cite{footnotechi}
giving a Wilson ratio $\chi_s/\gamma=$ 1.29 to 1.34$R_0$.
 For  YBa$_2$Cu$_3$O$_7$, where there is some uncertainty from the Cu-O chain contribution, and in the precise value of
 $\gamma(300)$,
  $\chi_s/\gamma\simeq1.2 R_0$~\cite{CooperLoramRev1996,LoramJPCS2001}.
So  the experimental data is closer to $R_0$  rather than the Wilson value of 2$R_0$. This is not a limitation  because
 Fig.~10.9 of Ref.~\cite{Hewson} shows that
for heavy fermion compounds such as CeCu$_2$Si$_2$, CeCu$_6$ and UPt$_3$,  $\chi_s/\gamma\simeq R_0$,
and as explained later, a ratio of 1.3 is consistent with the present picture.
 Numerical treatments of the $t-J$ model~\cite{Prelovsek2000},  derived from the 2D Hubbard model,  give
 $\chi_s/\gamma\simeq R_0$ for $0.1<p<0.2$  and $T<J/k_B$.   Large values of $T_K$ do occur for dilute alloys, e.g.
 \underline{Al}Mn where $T_K$ = 600~K at
low $T$ falling to 470~K near 300~K because of thermal expansion and the strong volume dependence of $T_K$~\cite{Miljak1976}. This  reduces
the $T$-dependence of $\chi_s(T)$ from the  $1/(T+T_K)$ law usually found for  alloys with lower $T_K$ and less thermal expansion.

 Here  we consider three localized  states, with occupancy 0, 1 and 2 described
  by the asymmetric single orbital Anderson model.
 Figs.~\ref{Fig1Costi}(b) and (c), taken from Fig.~5 of~\cite{Costi1994}, show the  spectral density for an isolated ion
 obtained by numerical renormalization
 group~(NRG) calculations
 for  $E_0=-U/2$, the symmetric case, and   various asymmetric cases,
 $-E_0/\Delta$ = 4, 3 and 2.   $|E_0|$ and $U-|E_0| >0$ are the  energies required to transfer an electron
from the singly-occupied state to the Fermi level of the conduction electrons at $E_F$, or from $E_F$ to the
 doubly-occupied state respectively.
 $U$, set equal to 4$ \pi \Delta$ in~\cite{Costi1994}, is the on-site $d-d$
Coulomb repulsion, $\Delta = \pi V^2 N(E)$~\cite{Anderson1961},
where  $V$ is the hybridization energy between the $d$ state and the conduction electrons and $N(E)$ is their DOS
for a given spin direction at the energy of the virtual bound state. The familiar triple peak structure in Fig.~\ref{Fig1Costi}(b)  has  two side peaks arising
  from the localized level and a
     peak near $E_F$
 from the Kondo resonance.
The weak $p$ dependence of
$\gamma$ and hence $T_K$ for OD Bi2212 and
 other OD cuprates  implies that   $E_0$ and
$\Delta$  are also only weakly dependent on $p$.

As originally shown by Kondo~\cite{Kondo1964} the central peak is caused by
  the $af$
 exchange interaction -$2J_{imp}\underline{S}_{imp}.\underline{s}_e$, with $J_{imp}<0$, between the spin on the magnetic ion, $\underline{S}_{imp}$ and
 the spin
  of a conduction electron $\underline{s}_e$.  The ${S}^+_{imp}$ and ${S}^-_{imp}$ operators
   do not commute, and because of this, higher order scattering
  processes contain the Fermi function for the conduction electrons. This leads to the logarithmic Kondo divergence in the  scattering rate,
   responsible for the
  resistance minimum observed in dilute magnetic alloys and for the formation of the Kondo resonance at $E_F$. It is a many-body effect
  because it depends on  the occupancy
   of many other conduction electron states.

The Kondo resonance has the same spatial symmetry as the
   electronic state of the localized level~\cite{Gruner1974,Blandin1968}.
This is why
 band-structures  of heavy fermion materials, calculated using
relatively standard methods, agree with the extremal orbits observed in quantum oscillation experiments, e.g.~\cite{Flouquet2017},
 because the band
structures  depend on this symmetry.  But the effective masses ($m^*$)  are often very large, because they are determined by the energy width of
 the Kondo resonance. For the cuprates this implies that the Fermi
surface is large, with the shape predicted by standard band theory,  but $m^*$ is larger.

 The effects of an applied magnetic field~($H$) and,
more generally, an exchange field, are  important here. In dilute Kondo alloys,
 $H$ suppresses the Kondo effect
because spin-flip scattering processes  become inelastic in a magnetic field~\cite{Gruner1974}. Early tunnelling studies
 where a fraction of a monolayer of Fe atoms was evaporated
on the oxide layer of an Al-Al$_2$O$_3$-Al junction~\cite{Bermon1978} showed evidence for this ``hole in the DOS'' and
 it is indeed ``non-states-conserving'', see Fig.~9  of~\cite{Bermon1978} shown in Fig.~5(SM)~\cite{Suppmat}. Experimentally~\cite{Monod1967,Rohrer1969} and
  theoretically~\cite{Horvatic1984,CostiMR2000}
 this
requires $\mu_BH \gtrsim k_BT_K$. A problem here is that such a large value of $H$,  $\approx k_BT_K/\mu_B$,
  is  incompatible with the appropriate condition  $ E_G\simeq \mu_BH$  when  $p\lesssim p_{crit}$ and $E_G$ is small.

This difficulty  is  absent for a molecular field in NRG calculations~\cite{Costi2007} for  the two-band Hubbard model plus a ferromagnetic, predominantly Ising,
 exchange interaction~$-2J_{ex}S^z_i.s^z$ between the  localized magnetic moments and the conduction electrons.  Note that these
  calculations refer to a concentrated heavy fermion-like
  spin system.  For $J_{ex}=0$ there are
  two side peaks  from the narrow band, i.e. the localized
magnetic states, and a
  central peak  from the broad band, equivalent  to the conduction electron band here.  But as shown
  in \cite{Costi2007} for large enough $J_{ex}$
  there is a quadruple peak
 in the
  DOS curves, i.e. an extra dip  at $E_F$, whose width is $\simeq 2J_{ex}$. It is possible that $J_{ex}$
   suppresses higher order spin-flip scattering processes, thereby
  suppressing the DOS at $E_F$. There appears to be no
  limitation regarding the ratio of $J_{ex}/k_BT_K$, in contrast to the effect of $H$ on Kondo alloys.

\begin{figure}
\includegraphics[width=75mm,height=70mm]{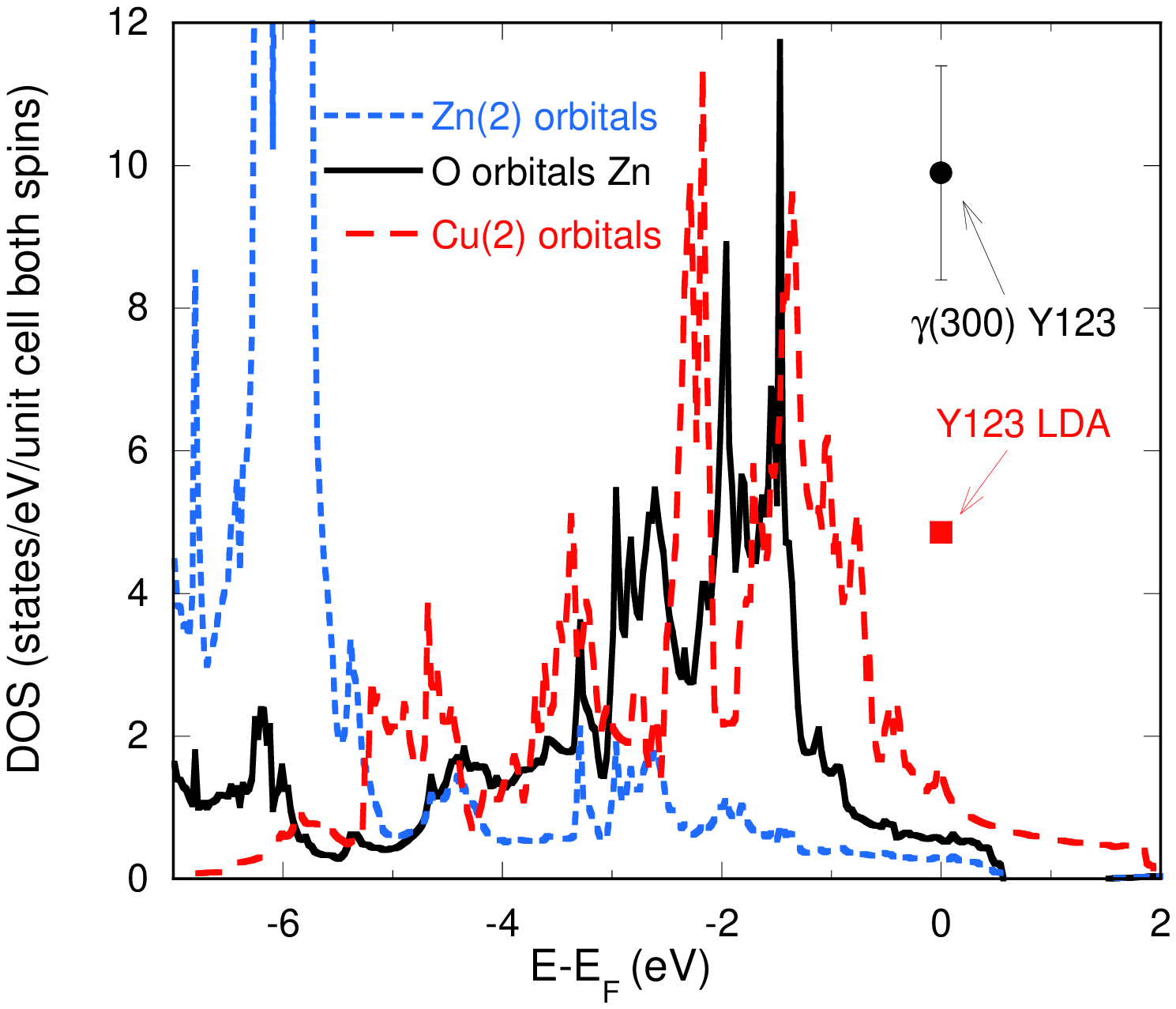}
\caption{Color online.
   Partial DOS $vs.$ energy $E-E_F$; in-plane oxygen orbitals of YBa$_2$Zn$_3$O$_7$, solid black line,
 $d$ orbitals  for Zn(2) atoms in YBa$_2$Zn$_3$O$_7$ and Cu(2) atoms in YBa$_2$Cu$_3$O$_7$,  short  and long dashed lines respectively.
  Points show
 $N(E_F)$ of YBa$_2$Cu$_3$O$_7$ from LDA  and   from $\gamma$(300). When expressed as states/eV/Cu atom (both spins), $\gamma$(300)
for YBa$_2$Cu$_3$O$_7$, gives 3.3$\pm$0.5, the same within errors as  2.9$\pm$0.2 states/eV/Cu atom (both spins)  from $\gamma$(200) of Bi2212.} \label{Fig2YBCOZn}
\end{figure}
 It is difficult to calculate $\Delta$ from first principles, even for noble-metal  hosts, see~\cite{Hewson}, Chapt.~9.7.   We attempted this by  calculating the
band-structure of
  the hypothetical compound
 YBa$_2$Zn$_3$O$_7$, in which all lattice parameters are kept the
  same as for YBa$_2$Cu$_3$O$_7$, using  the Wien2K code~\cite{Wien2K}. In a sense this is equivalent to
   experimentalists using Lu compounds with a
  full $f-$shell as a reference for Yb compounds.  The Brillouin zone is sketched in  Fig.~6(SM)(a)~\cite{Suppmat} and dispersion curves shown in
  Fig.~6(SM)(b)~\cite{Suppmat}.
The Zn $d$-levels  are well below  $E_F$ as  shown  by the
  the DOS $vs.$ energy plots in Fig.~\ref{Fig2YBCOZn} and there are approximately 10 $d$-states per Zn atom, i.e. a full $d$ shell
  as expected from chemical arguments.   In-plane states near
    $E_F$ do have a small amount of $d$ character but as shown in Fig.~\ref{Fig2YBCOZn} this is only a few $\%$ of the DOS
     derived from the measured value of $\gamma$.
    In contrast,  band structure calculations for YBa$_2$Cu$_3$O$_7$ using the local density approximation~(LDA)
 give substantial contributions to $N(E_F)$ from the Cu $3d$  orbitals  forming the Cu-O$_2$ plane bands. The full LDA DOS for  YBa$_2$Cu$_3$O$_7$ (not shown)
    agrees well with earlier work~\cite{Pickett1989}  and
  $\gamma$(300) is only moderately enhanced over the LDA value, see Fig.~\ref{Fig2YBCOZn}.
  But $N(E_F)$ from the in-plane oxygen bands and the in-plane Zn atoms in YBa$_2$Zn$_3$O$_7$
    is much smaller, so in the present picture the measured values of $\gamma$ for  YBCO and Bi2212 arise from the
     Kondo effect - Eqn.~\ref{gamma}.
 Fig.~6(SM)(c)~\cite{Suppmat} shows  the Fermi surface of YBa$_2$Zn$_3$O$_7$, with two  quasi-cylindrical sheets of
   similar size  from the bands crossing $E_F$ between S and X and  S and Y,
 and an open surface,   from the band crossing
   $E_F$ between Y and G. From Fig.~6(SM)(b)~\cite{Suppmat} the average value of $E_F$ for
   the two  quasi-cylindrical  bands of the  Zn compound
   is 0.54~eV relative to the top of the
   band. The
  average value of~$k_F$ in the S-X-Y plane is 0.35~$\pi/a$ where $a$ = 0.382~nm is the in-plane lattice spacing,
giving  $p$  for the two cylindrical bands which agrees well with the more
 precise value obtained by integrating the   DOS plots in Fig.~\ref{Fig2YBCOZn}, for the in-plane  O(2), O(3) and Zn(2) atoms.  Namely
 $p$= 2 x 0.19/unit cell, in surprisingly good agreement with empirical experimental estimates for
  YBa$_2$Cu$_3$O$_7$~\cite{Presland1991,Liang2006,Obertelli1992}.
     It would be interesting to see whether this procedure works for
 the two-chain compound YBa$_2$Cu$_4$O$_8$.

The complex band structure below  $\approx$ -1~$eV$ shown in Figs.~\ref{Fig2YBCOZn} and~6(SM)(b)~\cite{Suppmat} prevents
us estimating
 $\Delta$ from first principles, so initially we set
  $\Delta$ = 1 $eV$, a typical value for 3$d$ impurities in noble
  metals~\cite{Anderson1961,Gruner1974}.  But, as explained in more detail in~\cite{Suppmat}, we can account for the experimental value of the Wilson ratio
  if $E_0/\Delta$ = -2.3, because for this value, linear interpolation of the data in Table 1 of~\cite{Costi1994}  gives the occupancy
   of the localized level, $n_0$ = 0.81 and an effective moment of 0.81$\mu_B$. This reduces the susceptibility by 0.81$^2$ and the Wilson ratio from
   the value 2 expected for an ion with $S$=1/2 to 1.31, in good agreement with experiment.
Theoretical  $S_e(T)$ curves were obtained by integrating
  the  specific heat data in Fig.~4 of ~\cite{Costi1994} divided by $T$. For $T_K$ = 800~K, $\gamma$ increases by a factor 1.19 between $T$= 200~K and $T$=0,
so the initial estimate
  of $T_K$ = 800~K  from $\gamma(200)$ of Bi2212 is reduced  to 672~K.
Theoretical curves
  for representative samples of Bi2212, one with no PG, $p$=0.19 and one with a PG, are linear over the measured range
   of $T\leq$220~K as shown in Fig.~7(SM)~\cite{Suppmat}.
In contrast those for  YBa$_2$Cu$_3$O$_{6.97}$ with no PG and  YBa$_2$Cu$_3$O$_{6.48}$ with $E_G/k_B$ = 320~K, in Fig.~8(SM)~\cite{Suppmat}, both having $T_K$=775~K,
show
deviations from linearity above 200~K. But these do not rule out the present picture because they could  arise from a modest
 decrease in $T_K$ caused by thermal expansion, or because the Cu atoms in the CuO chains and the CuO$_2$ planes are inequivalent.

 Hence for Bi2212 $T_K$=672~K
and for $E_0/\Delta$ = -2.3, $k_BT_K/\Delta$ = 0.073, and $\Delta$= 0.79~$eV$. The observation of significant $g-$factor anisotropy
 in YBCO~\cite{Kokanovic2016} and references therein, supports the present localized picture. Here the concept of valency is meaningful, the
   Cu$^{2+}$ ($d^9$) state will be dominant and for $-E_0= 2.3\Delta$ there is a certain admixture of Cu$^{3+}$
($d^8$). The $d^{10}$ and $d^9$ states are  analogous to those in heavy fermion Yb compounds with 0 or 1 hole in the 4$f$ shell, which
 implies that all
the Cu $d$ electrons are contributing to $U$.
  As shown in Fig.~\ref{Fig1Costi}(c) the half-width-half-maximum (HWHM) values of the spectral density at $\omega<0$
  for $-E_0/\Delta$ = 3 and 2 are $\simeq$0.076$\Delta$ and 0.135$\Delta$ respectively giving 0.117$\Delta$ for $-E_0/\Delta$ = 2.3. The carrier density
    in the resonance $n_K$,
  obtained by assuming that it has a Lorentzian shape for $\omega<0$ and normalizing to the area of the lowest broad peak in
  Fig.~\ref{Fig1Costi}(b) that contains $\simeq$~1 electron, agrees well
with (1-$n_0$)/2 = 0.095. For $\Delta$ = 0.79~$eV$ the HWHM is $T_L$=1072~K.
  When transformed in the usual way into $\underline{k}-$ and then by Fourier transform into
$\underline{r}-$ space, \emph{and in the absence of a PG}, the carrier density  decays as $n_K\exp(-r/R)$
  with a range $R=\frac{\hbar v_F}{k_BT_L}$.
 The resonance is centered at the Fermi level and
because of the sharpness of the Fermi surface, the Fourier transform could also give rise to  Friedel or Rudermann-Kittel-Kasuya-Yosida type oscillations in
  $n_k(r)$, that could also be important but are not considered here.
The  Fermi velocity $v_F$= 1.6~10$^7$ cm/sec, estimated for a cylindrical  surface containing 1+$p$ holes with
 $m^*=5.2m_e$~\cite{Bangura2010}, agrees well with nodal ARPES data  for
UD and optimally-doped (OP) crystals of   the single layer cuprate Hg1201 shown in Fig.~2(b)
   of~\cite{Vishik2020},  1.7, 2.2 and 2.0~10$^7$ cm/sec for UD70, UD80 and OP98.
 This agreement supports our statement
  that $T_K$ and hence $m^*$ vary little with $p$.

The above values of $v_F$ and $T_L$ give $R=2.83a$.
 From the previous spatial symmetry argument, $n_K$ will have $|d_{x^2-y^2}|$ symmetry and be larger
  in the (0,1) and (1,0) directions, where $af$ spin fluctuations cause neighboring
  spins to be anti-parallel over a certain spatial range, $\xi$.  A typical value for $p$ = 0.1 is $\xi$ = 0.9 nm, i.e.
  2-3$a$~\cite{Rossat1992}.  If $\xi=2a$ then only the 4 nearest neighbors have correlated spins. Their $af$ Kondo screening
   ``clouds''  will give a $\emph{ferromagnetic}$
  conduction electron spin polarization
  at the central site. Including the effect of the PG on both $n_k$ and $R$ (by self-consistently representing  the PG by a narrower
  negative
  Lorentzian) reduces $n_k$ at the central site by a factor $\simeq 2$, giving a magnetic moment from
  the 4 overlapping Kondo ``clouds'' of 2 x 0.095 x $\exp(-1/2.83) \mu_B$ = 0.133$\mu _B$.
   It is difficult to prove that this spin polarization
  is equivalent to $J_{ex}$ in~\cite{Costi2007}, but, for $\chi_s$ = 2.3 10$^{-4}$ emu/mole-Bi2212,
   a spin polarization of 0.133 $\mu_B$/Cu is produced by a field $H$ where $\mu_BH/k_B$ = 433~K, so  $J_{ex}/k_B$ = 866~K.
   From Fig.~2 of~\cite{Costi2007} the
    half energy gap, equivalent to $E_G$ here, is  $\simeq J_{ex}$, in reasonable agreement  with the experimental
   value $E_G/k_B\simeq$600~K for $p$ = 0.1.
   Furthermore, the imaginary part of the self energy
   is unusual and typical of a ``bad metal''~\cite{Costi2007} in agreement with experiments on the cuprates.  However, more work is
  needed to see whether $J_{ex}$ falls linearly with $p$ and whether calculations allowing for   asymmetry in the DOS about $E_F$ on a scale
   of $J_{ex}$, i.e. $E_G$,
 would account    for the strong $p$-dependence of the TEP and its scaling with
   $k_BT/E_G$~\cite{CooperLoramRev1996,Cooper2000}, shown in Fig.~9(SM)~\cite{Suppmat}.

Alternatively, in a related picture,  the effect of the $af$ fluctuations, i.e. the interactions between neighboring Cu spins, could be considered directly, without
appealing to
 the work of Ref.~\cite{Costi2007}. These will reduce the entropy of the Cu spins and hence that available for the Kondo resonance.
Key theoretical questions are (a) can they cause
 a deep PG even when the interaction energy ($W$) is much smaller than $k_BT_K$ and (b) does $W$ fall linearly with $p$ and become very small
 for $p\geq p_{crit}$?  Experimental evidence  for (b) is given by ratio of
   the $^{63}$Cu and $^{17}$O NMR relaxation rates $T_1^{-1}$~\cite{TallonLoram2001} which is a measure of the strength of low frequency spin
   fluctuations for which the effect of the PG on $\chi_s$ cancels out.
 It does indeed fall linearly with $p$, approaching zero  at $p_{crit}$. Analysis~\cite{Tallon2008} of neutron scattering data shows
 the same decrease for
 $af$ spin fluctuations of energy below 50~$meV$, so on the time scale of the Kondo resonance, with   $k_BT_K \approx$ 60~$meV$,
  they will provide a  molecular field that is nearly static.

 Within the usual
    localized picture $W$ arises from superexchange via
   completely occupied or completely empty oxygen $2p$ states.
For a  band with Fermi energy $E_F$, we should consider  Wannier functions that have
 a  lifetime of order  $\hbar/E_F$. $E_F$ will increase with
$p$, thereby reducing $W$, which could account for point (b) above.
 As sketched for example in Fig.~13 of~\cite{CooperLoramRev1996}, only those parts of the large Fermi surface spanned by the
 $af$ wave vector $\underline{Q}\approx(\frac{\pi}{a},\frac{\pi}{a})$, will be affected by $af$ fluctuations.  This accounts for the  Fermi
  arcs, where  $E_G$ =0,
seen by ARPES~\cite{Chen2019} and scanning tunnelling microscopy (STM)~\cite{Davis2014}.  It also gives a
 strong energy dependence of the scattering rate
 responsible for the TEP~\cite{CooperLoramRev1996}, see also~\cite{Hildebrand1997}, and its
 scaling with $k_BT/E_G$ shown in ~\cite{CooperLoramRev1996} and  Fig.~9(SM)~\cite{Cooper2000,Suppmat}.

Evidence for an energy scale $\approx$60$~meV$ in the many spectroscopic studies of cuprates above $T_c$ would be a good experimental test of the present picture.
 ARPES data do show structure near 50~$meV$, e.g.~\cite{Vishik2020} that is presently ascribed to other factors.
However, we are dealing  with a many-body effect, so  comparison of  ARPES, STM~\cite{Davis2014} and optical
 data ~\cite{Basov_Timusk2005,Tajima2016} for cuprates with those for heavy fermion compounds as well as with theory could be informative.
 Another test would the apparent correlation between
 the $T^1$ term in the
 electrical resistivity and the superfluid density~\cite{Culo2021} in OD compoumds.
One prediction of the present approach is that $m^*$ and related properties of OD Tl2201 crystals could be strongly pressure dependent
  since for classical Kondo alloys
 the volume~($V$) dependence of $T_K$ is large, -d$\ln T_K$/d$\ln V$ = 16-18~\cite{Miljak1976,Schilling1973}.

 In summary, guided by the unusual behavior of the electronic entropy  revealed by  specific heat
  measurements~\cite{LoramJPCS1998, LoramJPCS2001, Loram1993}, we propose that
  the pseudogap is  the energy scale over which a Kondo-like enhanced
 DOS at the Fermi energy is suppressed by $af$ spin interactions.

The author would like to thank A. Carrington, who also calculated the band structure of  YBa$_2$Zn$_3$O$_7$,
 J. L. Tallon and V. Zlati\'{c} for helpful suggestions and
discussions  and to acknowledge a long and happy collaboration with the late Dr. J. W. Loram.

\end{document}